\documentclass[aps,twocolumn,superscriptaddress,nofootinbib,10pt,showkeys]{revtex4}
\usepackage{hyperref}
\usepackage[english]{babel}
\usepackage{setspace}
\usepackage{amsfonts,amsmath,amssymb}
\usepackage{graphics}
\usepackage{multirow}
\usepackage{booktabs}
\usepackage{xcolor}

\newtheorem{Lemma}{Lemma}

\newcommand{\ie}{{\it i.e.}~}
\newcommand{\eg}{{\it e.g.}~}

\newcommand{\be}{\begin{equation}}
\newcommand{\ee}{\end{equation}}
\newcommand{\ba}{\begin{eqnarray}}
\newcommand{\ea}{\end{eqnarray}}
\newcommand{\bmt}{\begin{multline}}
\newcommand{\emt}{\end{multline}}
\newcommand{\fd}{\left(1+2X B/A\right)}
\newcommand{\fdB}{\left(1+2X B\right)}
\newcommand{\mtc}[1]{\mathcal #1}
\newcommand{\st}{Scalar-Tensor theories }

\keywords{Horndeski Theory, Scalar-Tensor gravity}

\begin{document}

\begin{abstract}

The Horndeski action is the most general one involving a metric and a scalar field that leads to second order field equations in four dimensions.
Being the natural extension of the well known Scalar-Tensor theories, its structure and properties are worth analyzing along the experience accumulated in the latter context. Here we argue that disformal transformations play, for the Horndeski theory, a similar role to that of conformal transformations for Scalar-Tensor theories {\em a l\`a} Brans-Dicke. We identify the most general transformation preserving second order field equations and discuss the issue of viable frames for this kind of theories, in particular the possibility to cast the action in the so called Einstein frame. Interestingly, we find that only for a subset of Horndeski Lagrangian such frame exists. Finally, we investigate the transformation properties of such frames under field redefinitions and frame transformations and their reciprocal relationship.
\end{abstract}

\title{Disformal invariance of second order scalar-tensor theories:\\
framing the Horndeski action}
\author{Dario Bettoni}
\email[E-mail: ]{bettoni@sissa.it}
\affiliation{SISSA/ISAS,
Via Bonomea 265, 34136, Trieste, Italy}
\affiliation{INFN, Sezione di Trieste, 
Via Valerio, 2, 34127, Trieste, Italy}

\author{Stefano Liberati}
\email[E-mail: ]{liberati@sissa.it}
\affiliation{SISSA/ISAS,
Via Bonomea 265, 34136, Trieste, Italy}
\affiliation{INFN, Sezione di Trieste, 
Via Valerio, 2, 34127, Trieste, Italy}

\date{\today}
\maketitle

\section{Introduction}
\label{intro}

There are very few theories that can rival  General Relativity (GR) for elegance, simplicity and longevity. This pillar of modern physics, thanks to its remarkable agreement with experiments, is nowadays unanimously considered as the standard model of classical gravitational interactions~\cite{Will:2005va,Turnyshev2009}. It might seems then a preposterous attitude the growing attention that in the last years has been devoted to generalised theories of gravitation. There are however very good reasons to not be satisfied with the present state of affairs. 

On the theoretical side, GR remains poorly understood in its foundations: we can construct very many alternative theories of gravitation but we do lack an axiomatic derivation of such theories and hence an authentic understanding of their reciprocal relation. Moreover, generalised theories of gravitation can be as well considered as different effective actions induced by physics beyond the Planck energy and as such their study as alternative models of gravitation could provide some insight on the long standing problem of building a quantum gravity theory. 

On the experimental side, we do lack  severe experimental constraints on GR from galactic scales upwards. Of course, we do know from both cosmology and astrophysics that GR plus a cosmological constant so far provides a very good description of the observed universe~\cite{Planck}. However, this comes at the price of accepting that $95\%$ of the energy/matter content of the universe is of unknown nature. Indeed, dark matter and dark energy have been among the most pressing motivations for the recent outburst of attention toward alternative theories. To name a few, $f(R)$ theories of gravity \cite{Sotiriou2010}, generalise Scalar-Tensor theories \cite{Fujii:2003pa}, TeVeS \cite{Moffat2006} and MOND \cite{Milgrom,Milgrom2009,Bekenstein2009}  (see  also \cite{Clifton2012} and references therein for a thorough presentation).

In particular, generalised Brans-Dicke (BD) Scalar-Tensor theories have acquired, since their initial proposal more than half a century ago~\cite{BD1961}, a most relevant role as the standard alternative theories of gravitation. The investigation of formal aspects of these theories has played a fundamental role 
for several theoretical and observational issues in gravitation. In particular, scalar-tensor theories have represented an ideal setting for understanding the thorny issue of the different representations of a given gravitational theory. For example, it has been realised that a whole class of higher curvature theories, $f(R)$ theories, can be recast as special cases of Scalar-Tensor theories (with the number of scalars related to the order of the initial field equations). Even more interestingly, the invariance of the action of generalised Scalar-Tensor theories under metric conformal transformations and redefinitions of the scalar field, can be used to relate several equivalent frames, for example trading off a space-time varying gravitational constant (i.e.~a non minimal coupling) for an GR-like gravitational sector (i.e~minimally coupled) associated to a matter action with field-dependent mass and coupling constants. 

It is worth stressing that such features are not only theoretically interesting, but are also relevant for the actual observational tests of the theory. So much so that the question of whether conformally related frames are physically distinguishable is still an open issue in the literature (see e.g.~\cite{Faraoni2}). Furthermore, this kind of investigations become even more important as one moves further away from GR into more general theories. 

Further generalisations of the \st theories have been extensively  investigated in the contexts of cosmology \cite{Faraoni}, Dark Energy (DE)\cite{AmendolaBook,Gubitosi2011,Gleyzes:2013ooa}
 and Inflationary models \cite{deFeliceinflation,Germani2011}, and have indeed provided very efficient frameworks for explaining (in a alternative ways w.r.t. GR) the observed properties of the universe.  

An extension of the \st framework that that has attracted a lot of interests is represented by the Horndeski action \cite{Horndeski1974}, recently rediscovered in the context of the Covariant Galileon theory \cite{Deffayet2011}. This action provides the most general Lagrangian for a metric and a scalar field that gives second order field equations and as such is a well motivated effective field theory. This class of \st as been extensively investigated since it includes, as sub-cases, basically all known models of DE and single scalar field inflation. However, this generality comes at a dear price. In fact, the physics derived from the full action is rather obscure and the theory has been investigated only in few regimes or for particular models, like the FLRW universe, so that a systematic investigation is still missing (see however~\cite{Mota2013,Amendola2013-2} for a first attempt in this direction and \cite{Amendola2012,Amendola2013} for a method to derive constraints in the context of DE models). 

Given the above mentioned fruitful interplay between \st and conformal transformations one may wonder whether a generalisation along this line might help shedding some light on the properties and structure of the Horndeski theory. This is the main motivation of the present work.
As we shall see in what follows, simple conformal transformations are not enough for this task, due to the more complicate structure of the Horndeski actions, and the use of generalised metric transformation will be required.  

An example of such generalised metric transformation is given by disformally related metrics. These have been proposed in \cite{Bekenstein1993} and applied first in the context of relativistic extensions of MOND-like theories \cite{Bekenstein1994} in order to account for measured light deflection by galaxies. Later they found applications in varying speed of light models \cite{Bassett2000}, dark energy models \cite{Zumalacarregui2010,Koivisto2012,Mota2013,defelice2012}, inflation \cite{Kaloper} and modified dark matter models \cite{Bettoni2011,Bettoni2012}. More recently, empirical tests of these ideas have been proposed in laboratories experiments \cite{Brax2012} as well as in cosmological observations \cite{vandeBruck:2012vq,vandeBruck2013,Brax2013}, signaling the important role that disformal transformation are playing in contemporary cosmology and gravitation theory. 

The paper is organized as follows. In section \ref{HD} we briefly introduce the Horndeski action and the disformal transformations, discussing the most general set of transformations that leave the action invariant. In section \ref{special} we discuss some specific cases of disformal transformations and identify the subset of Horndeski actions admitting an Einstein frame. In section \ref{Frames} we discuss in more detail the issue of disformal frames and their properties
under field redefinitions and metric transformations, providing some explicit examples and discussing their reciprocal relationship. Finally, in section \ref{conclusions} we draw our conclusions.

\section{The Horndeski action and disformal transformations}
\label{HD}

The Horndeski Lagrangian \cite{Horndeski1974} is the most general Lagrangian that involves a metric and a scalar field that gives second order field equations in both fields in four dimensions. Recently generalised to arbitrary dimensions by Deffayet et al. in \cite{Deffayet2011}, it is the natural extension of \st {\em a l\`a} Brans--Dicke. 

Horndeski theory remained a sort of theoretical curiosity for more than thirty years but it was recently rediscovered as a powerful tool in cosmology. In fact, its generality (within the bound of second order field equations) made of it an ideal meta-theory for scalar-tensor models of dark energy and dark matter. However, up to now, no structural analysis analogue to the one carried out for standard \st was performed. In particular there is no obvious extension of the concept of equivalent frames and no principle to fix the shape of the free parameter functions. In order to address this points, after briefly reviewing the Horndeski action and disformal transformations, we shall discuss here the behaviour of this theory under such extended class of metric transformations.

\subsection{Horndeski Lagrangian} 

The Horndeski action, rephrased in the modern language of Galileons \cite{Kobayashi2011}\footnote{Notice that we have a different sign convention w.r.t. \cite{Kobayashi2011} due to the different definition of the function $X\equiv \nabla_{\mu}\phi\nabla^{\mu}\phi/2$} can be written as follows 
\be
\mathcal{L}=\sum_i\mathcal{L}_i\,,
\label{Horndeski}
\ee
where
\ba
\mathcal{L}_2&=&K(\phi,X)\,,
\\
\mathcal{L}_3&=&G_3(\phi,X)\square\phi\,,
\\
\nonumber
\mathcal{L}_4&=&G_4(\phi,X)R+\\
&&-G_{4,X}(\phi,X)\left[(\square\phi)^2-(\nabla_\mu\nabla_\nu\phi)^2\right]\,,
\\
\nonumber
\mathcal{L}_5&=&G_5(\phi,X)G_{\mu\nu}\nabla^\mu\nabla^\nu\phi+\\
\nonumber
&&\frac{G_{5,X}}{6}\left[(\square\phi)^3-3(\square\phi)(\nabla_\nu\nabla_\mu\phi)^2+2(\nabla_\mu\nabla_\nu\phi)^3\right]\,,\\
\label{Horndeskicoeff}
\ea
where $X=\nabla_{\mu}\phi\nabla^{\mu}\phi/2$, $(\nabla_{\mu}\nabla_{\nu}\phi)^{2}= \nabla_{\mu}\nabla_{\nu}\phi\nabla^{\mu}\nabla^{\nu}\phi$ and $(\nabla_\mu\nabla_\nu\phi)^3=\nabla_\nu\nabla_\mu\phi\nabla^\nu\nabla^\lambda\phi\nabla_\lambda\nabla_\mu\phi$ while $G_{i,X}=\partial G_{i}/\partial X$, R is the Ricci scalar and $G_{\mu\nu}$ is the Einstein tensor. The coefficient function $G_{4}$ has the dimensions of a mass square and it plays the role of a varying Gravitation constant, while $G_{5}$ has those of a mass to the fourth power. The field $\phi$ is taken to have mass dimension 1. 
As said, this gravitational action is the most general one that can be built with a metric and a scalar field, providing second order field equations in four dimensions. Notice that, despite the presence of second order derivatives in the action, no new degree of freedom is introduced, thus evading 
Ostrogradski's theorem, which states that such extra degrees of freedoms lead  to classical
instabilities \cite{Woodard:2006nt}. We will not discuss here the equations of motion referring the interested reader to \cite{Deffayet2011,DeFelice2011} for a general analysis.
Let us instead focus our attention on some important properties of this Lagrangian. 

First of all notice that, beyond the usual conformal non-minimal coupling, there is another source which couples the Einstein tensor to second order derivatives of the field. This represents a novelty as, contrarily to what happens for the coupling to the Ricci scalar, in this case we have a direction dependent coupling.
Secondly all the sub-Lagrangians give second order field equations independently so that one could in principle neglect some of them without spoiling the second order nature of the field equations. However, as is shown in appendix \ref{disftrans}, eventually neglected terms can alway be generated through redefinitions of the field variables. Finally, we notice that compared with the standard \st action the non minimal coupling (NMC) coefficients now depends also on the kinetic term. 

Given that this model is a generalisation of standard \st one may wonder whether suitable metric transformations can be introduced also in this case, leaving the action invariant and linking alternative frames. It not hard to realise that simple conformal transformations have limited power in this sense. In standard \st these transformations allow to replace by constants some of the field dependent coefficients. However, the various terms appearing in the Horndeski action ($K(\phi,X),G_i(\phi,X)$) are also dependent on the kinetic term $X$ and hence more general transformations are clearly needed. 

The most natural extension of the conformal transformation in the sense would be $A(\phi)\rightarrow A(\phi,X)$. However, even if this can remove the non-minimal coupling in the $\mtc{L}_{4}$, it is basically ineffective on the the non-minimally coupling provided by $\mtc{L}_{5}$. Moreover, this generalized conformal transformation contains derivatives of the field and hence one must be careful that those do not end up introducing higher derivatives in the equations of motion. In this sense, the next natural candidate for a suitable set of metric transformations is then represented by the disformal ones.

\subsection{Disformal transformations}  

Disformal transformations are defined by the following relation\footnote{More general formulations may be possible, for example including higher derivatives of the scalar field or by adding vector fields \cite{Bekenstein1993}.}
\be 
\bar{g}_{\mu\nu}=A(\phi,X)g_{\mu\nu}+B(\phi,X)\phi_{\mu}\phi_{\nu}\,,
\label{generaldisformal}
\ee
where the disformal functions $A$ and $B$ now depend on both the scalar field $\phi$ and its kinetic term $X$ and where we have defined for convenience $\phi_{\mu}=\nabla_{\mu}\phi$. We can classify the properties of this generalisation in two main categories: first the new functions do not simply depend on the local value of the field but also on the metric itself, hidden inside the definition of the kinetic term. Secondly we have a translation along the lines of variation of the field which means that the new metric will also depends on the way the field is changing through space-time. 

When dealing with metric transformations one has to ensure that the new metric is still a good one. We can formally define the goodness of a metric transformation with a set of properties: it must preserve Lorentzian signature, it must be causal and it has to be invertible, with a non zero volume element.
All these properties directly translate into constraints on the two free functions $A$ and $B$ which we are going to discuss one by one.
\begin{description}
\item[Lorentzian signature] Consider a frame in which $\phi_{\mu}=(\phi_{0},\vec 0)$. Then the Lorentzian requirement can be translated into\\
\be
\bar g_{00} = A(\phi,X)g_{00}+B(\phi,X)\phi_{0}\phi_{0}<0\,.
\label{lorentzian}
\ee
This constraint must hold true for all values of the field and its derivative. Given that we cannot exclude that for some values of the field variables the function $B$ can be zero, a first requirement is that $A>0$. This is the usual requirement made also for standard scalar-tensor theories. Then by multiplying equation \eqref{lorentzian} with $g^{00}$ we found that the condition to be fulfilled for preventing $\bar g_{00}$ from sign inversion is:
\be
A(\phi,X)+2B(\phi,X)X>0\,.
\ee
As a consequence, to have this relation to hold true for all values of $X$, it is necessary to have some kinetic dependence at least in one of the two disformal functions.
This result was first derived in \cite{Bruneton} (see also the original paper by Bekenstein \cite{Bekenstein1993}). However in \cite{Mota2013} it was argued that  the dynamics of the scalar field can be such that it is possible to keep the metric Lorentzian even with no $X$ dependences in the disformal functions $A$ and $B$. For example this can happen when the scalar fields enters a slow-roll phase \eg when thought to be the field responsible for Dark Energy. However this subject is not yet fully understood and, being not mandatory for our purposes its investigation, in the next we will assume that both metrics are Lorentzian for all the values of the scalar field and its kinetic term. 

\item[Causal behaviour] The disformal metric can have, depending on the sign of the $B$ function light cones wider or narrower than those of the metric $g$. This may lead to think that particles moving along one metric may show superluminal or a-causal behaviour. However, the requirement of the invariance of the squared line element and recalling that physical particles satisfies $ds^{2}<0$ is enough to ensure causal behaviour. This objection has been discussed in some details in \cite{Bruneton2007B}.

\item[Invertible]  We also must be sure that an inverse of the metric and the volume element are never singular. The inverse disformal metric is given by:
\be
\bar{g}^{\mu\nu}=\frac{1}{A(\phi,X)}g^{\mu\nu}-\frac{B(\phi,X)/A(\phi,X)}{A(\phi,X)+2B(\phi,X)X}\nabla^{\mu}\phi\nabla^{\nu}\phi\, ,
\ee
  while the volume element is given by $\sqrt{-\bar{g}}=A(\phi)^{2}\fd^{1/2}\sqrt{-g}$. The constraint derived from these requirements are weaker than those already obtained hence there are no new potential issues.

\end{description}

From this analysis we learn that the extension of conformal transformations to disformal ones is well posed, even if all previous points deserve a deeper analysis which, in any case, is beyond the scopes of the present paper and is left for further studies. 

Disformal metrics seem to be good candidates for our purposes as they possess, beyond a purely conformal term, another one which is a deformation of the metric along the direction of variation of the field and indeed disformal transformations have for the Horndeski action a role very similar to that of conformal transformations for standard Scalar-Tensor theories.

\subsection{Invariance of the Horndeski Lagrangian under disformal transformations}
\label{HornInv}

The ability of the Horndeski action to give second order field equations resides in a fine cancelation between higher derivatives coming from NMC terms and those produced from derivative counterterms, which requires the coefficient functions of the second field
derivatives in $\mathcal{L}_4$ and $\mathcal{L}_5$  to be proportional to the derivative of $G_i$ with respect to $X$ \cite{Deffayet:2009wt}.

It is easy to see that this requirements already reduces the freedom in the disformal functions $A$ and $B$. In fact, any kinetic dependence of these two terms would lead unavoidably to the breaking of the Horndeski structure, i.e.~to higher order equations of motion. We prove this through some examples in appendix (\ref{NoX}) while here we give a first principle argument why one should expect this to happen.
The ability of the Horndeski action to give second order field equations lies on the antisymmetric structure of second derivatives terms, as has been made clear in \cite{Deffayet2011}.  Consider the $\mtc{L}_{4}$ part of the Lagrangian. This can be rewritten in the following form:
\begin{eqnarray}
\mtc{L}_{4}&=&\left(g^{\mu\beta}g^{\nu\alpha}-g^{\mu\nu}g^{\alpha\beta}\right)\left[G_{4}(\phi,X)R_{\mu\nu\alpha\beta}\right.\\
&&\left.-G_{4,X}(\phi,X)\nabla_{\mu}\nabla_{\nu}\phi\nabla_{\alpha}\nabla_{\beta}\phi\right]
\label{exemple}\, ,
\end{eqnarray}
where the antisymmetric structure is made clear. Given that we have to preserve this structure in order to keep the equations of motion second order, any transformation operated on the fundamental variables $\varphi$ and $g_{\mu\nu}$ has to be necessarily reabsorbed into the coefficient function G4 and its derivative modulo a surface term. However, any kinetic dependence in the disformal functions will spoil this structure. In fact, consider the transformation property of the second derivatives of the scalar field under the conformal transformation $\hat g_{\mu\nu}=A(X)g_{\mu\nu}$
\begin{eqnarray}
\nonumber
\nabla_{\mu}\nabla_{\nu}\phi&\rightarrow& 
 \nabla_{\mu}\nabla_{\nu}\phi +\frac{A_{,X}}{A}\left[g_{\mu\nu}\phi^{\alpha}\phi^{\beta}\nabla_{\alpha}\nabla_{\beta}\phi+\right.\\
 &&-	\left. \phi_{\mu}\phi^{\alpha}\nabla_{\alpha}\nabla_{\nu}\phi-\phi_{\nu}\phi^{\alpha}\nabla_{\alpha}\nabla_{\mu}\phi\right]\, .
\end{eqnarray}
When inserted in (\ref{exemple}), among other terms, the following one is generated
\be
\sim 4G_{4,X}\left(\frac{A_{,X}}{A}\right)^{2}\phi^{\mu}\phi^{\nu}\phi^{\alpha}\phi^{\beta}\nabla_{\mu}\nabla_{\nu}\phi\nabla_{\alpha}\nabla_{\beta}\phi\, ,
\ee
which is clearly symmetric in the four indices and hence will produce higher than second derivatives in the equations of motion. One may wonder whether there may be counterterms coming from curvature that eliminate this but, as shown in the appendix, this is not the case.
This result may seem an artifact of the transformation used. In this sense, it has been recently shown how, in some specific cases, such higher than second derivatives can be eliminated in the equations of motion using hidden dynamical constraint \cite{Zumalacarregui:2013pma}. However, further analysis is required to see whether this result can be generalized to the full Horndeski action. 

We hence conclude that in order to  be sure to preserve second order field equations, we have to restrict our analysis to the following class of disformal transformations\footnote{Even though we do not provide a formal proof that this relation is the most general that leaves invariant the Horndeski action, we notice that more general transformations, despite possible, have to introduce higher derivatives of the scalar field \cite{Zumalacarregui:2013pma}.}
\be 
\bar{g}_{\mu\nu}=A(\phi)g_{\mu\nu}+B(\phi)\phi_{\mu}\phi_{\nu}\, .
\label{disformal}
\ee
In appendix \ref{disftrans} we show that this transformation preserves the antisymmetric structure of the Horndeski action as its effects happen to simply renormalize the coefficient functions $K$ and $G_{i}$s. We refer the reader to this appendix for the detailed transformation properties while here we discuss the meaning of this and analyze some relevant sub-cases.
As a concluding remark, let us add that the Horndeski action is clearly also invariant under the field rescaling $\phi\rightarrow s(\phi)\phi$ (this is explicitly discussed this in appendix \ref{fieldinvariance} where we consider the effects of this transformation on the Horndeski coefficient functions). This property will play an important role later on in our discussion when we shall deal with the equivalence of disformal frames.

\section{Special cases}
\label{special}

The structural invariance of the action under disformal transformations translates into the statement that such transformations represent a symmetry of the Horndeski action so that all the functions are defined modulo a conformal and a disformal transformation. This reminds very closely the case of standard \st where invariance under conformal transformations is used to reduce the number of free functions that defines the theory. 
However, the generalisation of this reasoning to the case of Horndeski theory is not straightforward. In fact, the subset of disformal transformations (\ref{disformal}) does not allow for kinetic term dependent coefficients, consequently one cannot  generically rescale the functions $(K(\phi,X),G_i(\phi,X)$ characterising the Horndeski action.   The next two subsection are devoted to analyse further this issue. We will study the transformation properties of the Horndeski action under pure conformal and disformal transformations separately and in particular we will provide the sub-class of Horndeski theories that admits a representation in which all NMC terms are eliminated via a disformal transformation.

\subsection{Purely conformal transformations}

Let us first consider the effects of conformal transformations $\bar g_{\mu\nu}=A(\phi)g_{\mu\nu}$ on the Horndeski action (\ref{Horndeski}), extending the well known results for these transformations in \st theories to this more general class of actions. The transformed Lagrangian coefficient functions read
\begin{multline}
\bar K(\phi,X)=A^{2}K(\phi,X_{C})+2XG_{3}AA'
+3X\frac{G_{4}A'\left[1-2A\right]}{A}\\	
+\frac{6G_{5}X^{2}A'}{A}\left[\frac{A''}{A}-\frac{A'^{2}}{A^{2}}\right]-2XH_{5,\phi}+
\frac{2G_{5,X}X^{3}}{A^{3}}A'^{2}\, ,	
\label{confK}
\end{multline}
\begin{multline}
\bar G_{3}(\phi,X)=AG_{3}(\phi,X_{C})-2G_{4,X}A' +
 X\left(-2H_{\square,\phi}\right. \\
  \left.-\frac{G_{5}A'^{2}}{2A^{2}}+\frac{2G_{5}A''}{A}+
\frac{G_{5,X}XA'^{2}}{A^{2}}\right)-H_{5}
\label{confG3}
\end{multline}
\be
\bar G_{4}(\phi,X)=A(\phi)G_{4}(\phi,X_{C}),\qquad
\bar G_{5}(\phi,X)=G_{5}(\phi,X_{C})\, ,
\label{confG45}
\ee
where
\be
X_{C}=X/A(\phi),\qquad H_{\square}=G_{5}\frac{A'}{A}\,, 
\label{eq:XC}
\ee
\be
H_{5}=\int dX\left[H_{\square,\phi}+\frac{G_{5}A''}{A}+\frac{5G_{5}}{2}\frac{A'^{2}}{A^{2}}+2G_{5,X}\frac{A'}{A}\right]\, .
\ee
The effects of the transformation are manifolds. 
First notice that there is a hierarchical propagation of terms from higher derivatives Lagrangians towards lower ones, that is to say $\mtc{L}_{i}$ generates terms that contribute to all $\mtc{L}_{j<i}$ so that even if in the original Lagrangian some terms were neglected they will inevitably appear after a conformal transformation. A special case is represented by the $\mtc{L}_{5}$ Lagrangian that cannot be generated this way.
Secondly, the conformal NMC $G_4(\phi,X)$, is modified by a multiplicative factor while the NMC with the Einstein tensor is unaffected apart from a redefinition of the kinetic term inside $G_{5}(\phi,X)$. 

Given that in general all the coefficient functions depend on both the scalar field and its kinetic term, it is clear that using only a conformal transformation we shall not be able to eliminate non minimal couplings for any choice of the conformal factor $A(\phi)$. Even in the special case when the coefficient functions depend only on the field and not on its derivatives, we are able to set at most $G_{4}(\phi)=1$ while retaining the generalised NMC between the Einstein tensor and the field derivatives (\ref{Horndeskicoeff}), given that part of the Lagrangian is not affected by conformal transformations, 
see Eq.~(\ref{confG45}). 

Notice that even if we were to take $G_{5}(\phi,X)$ to be a function of the scalar field only, we would not be able to eliminate it. In fact, we then have the following relation
\begin{multline}
G_{5}(\phi)G_{\mu\nu}\nabla^{\mu}\nabla^{\nu}\phi=G_{5,\phi}XR-G_{5,\phi}\left[(\square\phi)^{2}-(\nabla_{\mu}\nabla_{\nu}\phi)^{2}\right]\\
-G_{5,\phi\phi}\left[2X\square\phi-\phi^{\mu}\phi^{\nu}\nabla_{\mu}\nabla_{\nu}\phi\right]\, ,
\label{G5toG4}
\end{multline}
that shows how in this case $\mtc{L}_{5}$ is a contribution to the $\mtc{L}_{4}$ (as well as to $\mtc{L}_{3}$ and $\mtc{L}_{2}$) that depends explicitly on the kinetic term and cannot be eliminated by a simple conformal transformation. 

\subsection{Purely disformal transformations}

We turn now our attention to the case of a pure disformal transformation, \ie when the conformal factor $A(\phi)$ is set to one while the disformal function $B(\phi)$ is left unspecified. Given that we are mainly interested on the effects of transformations on the NMC terms we will report here only the relevant coefficient functions. The remaining ones can be easily derived from the equations in appendix \ref{disftrans} and the discussion of the effects of the disformal transformation on them is analogous to that for conformal transformation.
 
In the case under consideration we have that the transformed NMC coefficient functions read
\begin{multline}
\bar G_{4}(\phi,X)=\fdB^{1/2}G_{4}(\phi,X_{D})\\
+\frac{G_{5}(\phi,X_{D})B'(\phi)X^{2}}{\fdB^{3/2}}-H_{R,\phi}(\phi,X)X\,,
\label{G4puredisf}
\end{multline}
\be
\bar G_{5}(\phi,X)=\frac{G_{5}(\phi,X_{D})}{\fdB^{1/2}}+H_{R}(\phi,X)\,,
\label{eq:G5puredisf}
\ee
where 
\be
\label{XD-X}
X_{D}=X/(1+2BX), \quad H_{R}(\phi,X)=B\int dX \frac{G_{5}(\phi,X_{D})}{\fdB^{3/2}}\,.
\ee
 Here we notice that the effects of the disformal transformation are richer than those of the conformal one. In fact, besides a conformal modification of $G_{4}$ we have other contributions to $\mtc{L}_{4}$ and in this case $G_{5}$ is modified as well. In particular, the modified coefficient functions receive corrections that depends on the kinetic term but, as can be seen from equations \eqref{G4puredisf} and \eqref{eq:G5puredisf}, even in this case one cannot generically eliminate the NMC.

Let us focus on this last point and study which constraints can be imposed on the coefficient functions of the Hornedski action so to be able to eliminate all the NMC i.e.~to use the disformal transformation so to obtain $\bar G_{4}=1$ and $\bar G_{5}=0$. 
The latter condition is satisfied if
\begin{multline}
\frac{G_{5}(\phi,X_{D})}{\fdB^{1/2}}+B\int\frac{G_{5}(\phi,X_{D})}{\fdB^{3/2}}dX=0 \\
\Rightarrow \int dX\left[\frac{G_{5,X}(\phi,X_{D})}{\fdB^{1/2}}\right]=0\, .
\label{G5constraint}
\end{multline}
In general, if $G_{5}=G_{5}(\phi)$ the above constraint is  automatically satisfied. We cannot exclude the existence of other solutions in which an $X$ dependence is also allowed, for example if the integrand function is fast oscillating. However, these will depend on the specific model chosen and would need to be investigated case by case. Finally, notice that this constraint is not influenced by the freedom in rescaling the scalar field.

In order to have no conformal coupling to gravity we have to impose
\be
1=\fdB^{1/2}G_{4}(\phi,X_{D})+\frac{G_{5,\phi}(\phi,X_D)X}{\fdB^{1/2}}-\tilde G_{5}X\, ,
\ee
where
\be
\tilde G_{5}(\phi,X)=\int dX\frac{G_{5,X}(\phi,X_{D})}{(1+2BX)^{1/2}}\, .
\ee
This allows to find the form of the untransformed function $G_4$ in terms of the transformed variable $X_D$. Inverting relation \eqref{XD-X}\footnote{This can be always done as the Jacobian of the transformation $\frac{dX_D}{dX}=(1+2BX)^{-2}$ is never singular.} we can find the form of the function in terms of the untransformed variable $X$ that is needed to satisfy the requirement, namely
\be
G_{4}(\phi,X)=(1-2B(\phi)X)^{1/2}-G_{5,\phi}(\phi)X +\tilde G_{5,\phi}(\phi,X)X\, ,
\ee
with
\be
\tilde G_{5}(\phi,X)=\int dX (1-2BX)^{1/2}G_{5,X}(\phi,X)\, .
\ee
Given that we want both constraints to be satisfied at the same time, we have then 
\be
G_5=G(\phi) \quad \mbox{and} \quad G_{4}(\phi,X)=(1-2B(\phi)X)^{1/2}-G_{5,\phi}(\phi)X\, ,
\label{G4Einstein}
\ee
which fixes once for all the functional dependence of the $G_{4}(\phi,X)$ function on the kinetic term.

We conclude that the following Lagrangian
\begin{multline}
S_{NMC}=\int d^{4}x\sqrt{-g}\left[ G_{4}(\phi,X)R-\right.\\ 
\left.G_{4,X}\left[(\square\phi)^{2}-(\nabla_{\mu}\nabla_{\mu}\phi)^{2}\right]
+G_{5}(\phi)G_{\mu\nu}\nabla^{\mu}\nabla^{\nu}\phi\right]\, ,
\label{eq:finalSNMC}
\end{multline}
 where $G_{4}$ is given by (\ref{G4Einstein}), is the only one that admits a disformal map able to eliminate all the NMC terms in the context of Horndeski theory. 
 
However, it is worth noticing that inserting equations \eqref{G5toG4} and \eqref{G4Einstein} in \eqref{eq:finalSNMC} all the terms depending on $G_5(\phi)$ ends up cancelling. Hence, if the function $G_{5}$ depends only on the scalar field, we conclude that the existence of a disformal metric able to cancel all NMC {\em requires} the absence of $\mtc{L}_{5}$.\footnote{It may seem that a constant $G_{5}$ could be included without spoiling our request of no NMC. However in this case $\mtc{L}_{5}$ reduces to a surface term and hence does not contribute to the dynamics.} 
We are hence left with the following action
 \begin{multline}
S_{NMC}=\int d^{4}x\sqrt{-g}\left[G_{E}(\phi,X)R\right.\\
\left.-G_{E,X}\left[(\square\phi)^{2}-(\nabla_{\mu}\nabla_{\mu}\phi)^{2}\right]\right]\, ,
\label{eq:finalfinalNMC}
\end{multline}
where
\be
G_{E}=(1-2B(\phi)X)^{1/2}\, .
\ee
While we have here considered a special subset of the $\phi$-dependent disformal transformations \eqref{disformal} we can easily extend our conclusions to  transformations including a conformal factor $A(\phi)$. Indeed, in this case the most general action allowing for a full elimination of the NMC would be the same as equation \eqref{eq:finalfinalNMC} modulo a conformal rescaling of the $G_E(\phi,X)$ function.

As a final remark, it is perhaps worthy stressing that, as noted in~\cite{Mota2013}, the non-relativistic limit of the action~\eqref{eq:finalfinalNMC} corresponds to the  quartic covariant term of the Galileon action with the appropriate non-minimal coupling to yield second order field equations~\cite{Rham}. 

\section{Disformal frames}
\label{Frames}

The invariance of an action under metric transformations implies the possibility to fix some of the free functions characterising the theory, similarly to what is done when choosing a gauge. Consequently, the number of the independent functions is reduced. In our specific case the Horndeski action 
\eqref{Horndeskicoeff} is invariant under both purely conformal and disformal transformations. This freedom allows us to define an infinite set of equivalent frames defined by different fixings of two of the free functions in the action (see \cite{Sotiriou, Flanagan2004} for a similar reasoning in standard scalar-tensor theories). 

Among all these equivalent representations of the theory two are most relevant as they correspond to somewhat opposite situations:  the Einstein and Jordan frame. 
For the sake of clarity we provide here generalised definitions relevant for the Horndeski actions under consideration here.

\begin{description}
\item [Jordan Frame]
In the \textit{Jordan Frame} the Lagrangian of the gravitational sector includes a non-minimally coupled scalar field meanwhile all the matter fields follow the geodesics of the gravitational metric (the stress energy tensor of the matter fields is covariantly conserved w.r.t.~the gravitational metric).
\item[Einstein Frame]
In the \textit{Einstein Frame} the gravitational dynamics is described by the standard Einstein--Hilbert Lagrangian (plus possibly a cosmological constant). However, matter fields are coupled to the gravitational metric via some function of the scalar field and its derivatives. They hence move on geodesics that can be different from the one determined by the metric defining the Ricci scalar. Moreover, the gravitational equations in absence of matter do not reduce to $R=0$, as in GR, but in general will retain the scalar field as a possible source.
\end{description}

We now proceed to recall the issue of frames and their equivalence in standard \st and then extend this to the case of the Horndeski action.

\subsection{Scalar-Tensor Theories and conformal transformations}
\label{scalartensor}

Scalar-Tensor theories of gravity \cite{Fujii:2003pa,Faraoni,Clifton2012} represents one of the simplest and most studied extensions of GR in which a scalar degree of freedom is added to the Lagrangian besides the metric and matter fields.  A minimal prescription for generalising GR is to promote the gravitational constant to a scalar field which must be provided with its own dynamics in order to preserve diffeomorphism invariance. Furthermore, the Einstein Equivalence Principle (EEP) allows such scalar field to also mediate the coupling of the matter to the metric (albeit in an universal way). This reasoning then leads to the following action:
\begin{multline}
S=\int d^{4}x\left[G(\phi)R -\frac{f(\phi)}{2}\nabla_{\mu}\phi\nabla^{\mu}\phi-V(\phi)\right]\\
+S_{m}[e^{2\alpha(\phi)}g,\psi]\,,
\label{STA}
\end{multline}
where the four functions $G(\phi)$, $f(\phi)$, $V(\phi)$ and $\alpha(\phi)$ are general functions of their argument. We will not enter into the details of the applications of this theory, referring to the above cited papers and to references therein for details, but we will focus on some more formal properties of this action. 

First of all, the above mentioned free functions in the action are actually redundant for fixing a particular action~\cite{Flanagan2004,Sotiriou}. 
Indeed, the invariance of action \eqref{STA} under the conformal transformations $\bar g_{\mu\nu}=\Omega^{2}(\phi)g_{\mu\nu}$ and the scalar field redefinitions $ \bar \phi= F(\phi)$ imply the possibility to freely choose two out of the four functions. Hence, implementations of \eqref{STA} differing only for the fixing of two of the four coefficient functions are indeed just different representation of the same physical theory~\cite{Sotiriou}.

For this class of theories the Einstein frame is defined by the choice $G(\phi)=1$ and $f(\phi)=1$ so that gravity is described by the standard Einstein--Hilbert action, the scalar field has a canonical kinetic term while matter fields follows the geodesics of a physical metric conformally related to the gravitational one. The Jordan frame is instead obtained choosing $\alpha(\phi)=0$ and $A(\phi)=\phi$. In this case we have that all fields follows the same metric but now the scalar field is non-minimally coupled to curvature and it may possess a non-standard kinetic term. 
The fact that the above two frames are picked up from \eqref{STA} by just fixing two of the four coefficients function implies their mathematical equivalence (\ie a varying gravitational coupling in the Jordan frame is translated into field dependent matter masses and couplings when the action is in the Einstein frame). 

The lesson that we want to capture with this short introduction is that when dealing with generalized actions like  \eqref{STA} one has to pay attention to their symmetries in order to correctly identify the set of equivalent frames (\ie different representations of the same theory) which one can alternatively use for more conveniently dealing with different physical issues.

This considerations becomes even more important as further modifications of gravity are introduced and complicated terms are added. In what follows we shall first investigate the issue for that class of Horndeski actions admitting an Einstein frame (as this frame is often adopted for physical investigations). Later, we shall extend the discussion to more general actions.

\subsection{Horndeski action and the Einstein frame}
\label{framing}

In section (\ref{special}) we have derived the most general action in the Jordan frame for which all NMC can be eliminated via the disformal transformation (\ref{disformal}). However, the discussion of the possible equivalence of frames requires us to include also the action for matter fields with possible generalised coupling to the metric. This leads to the following completion of \eqref{eq:finalfinalNMC} 
 \begin{multline}
S=\int d^{4}x\sqrt{-g}\left[G(\phi,X)R+\right. \\
\left.-G,_{X}(\phi,X)\left[(\square\phi)^{2}-(\nabla_{\mu}\nabla_{\mu}\phi)^{2}\right]+\right. \\
 \left.+K(\phi,X)+G_{3}(\phi,X)\square\phi\right]+S_{m}[\bar g,\psi]\,,
\label{Einsteinfinal}
\end{multline}
where
\be
G(\phi,X)=C(\phi)^{2}\left(1-2\frac{D(\phi)}{C(\phi)}X\right)^{1/2}\,,
\label{G}
\ee
$S_{m}$ is the total matter action defined in terms of the physical metric 
\be
\bar g_{\mu \nu}=e^{\alpha(\phi)}g_{\mu\nu}+\beta(\phi)\phi_{\mu}\phi_{\nu}\,,
\label{eq:alphbet}
\ee
and $\psi$ stands generically for matter fields.

In the action appear six free functions, four related to the field-metric couplings, $C(\phi)$, $D(\phi)$, $\alpha(\phi)$ and $\beta(\phi)$ and two defining the minimally coupled scalar field Lagrangian, $K(\phi,X)$ and $G_{3}(\phi,X)$. Thanks to the invariance under both conformal and disformal transformations we can fix two out of the four metric functions $C(\phi)$, $D(\phi)$, $\alpha(\phi)$ and $\beta(\phi)$ with appropriate choice of the functions $A(\phi)$ and $B(\phi)$ appearing in the disformal transformation \eqref{disformal}. In principle, we could act on $K(\phi,X)$ and $G_{3}(\phi,X)$ but given their generic dependence on the kinetic term, \eqref{disformal} is not effective for fixing them. 
Hence, with a general disformal transformation \eqref{disformal}, we can define a Jordan and an Einstein frame in the same sense as it can be done for standard scalar-tensor theories. 

However, we can also use the invariance of the Horndeski action under field rescaling to further constrain the number of independent functions (as in the case of action \eqref{STA}). In fact, as shown in appendix \ref{fieldinvariance}, we can always rescale the field $\phi$ by a function, provided that this does not lead to a constant. This amounts to say that we can fix one more of the free functions $\alpha(\phi)$, $\beta(\phi)$, $C(\phi)$ and $D(\phi)$ to arbitrary values so that the Einstein and Jordan frames defined above represents a class of equivalent theories that can be further fixed with a field redefinition. 
We conclude that implementations of \eqref{Einsteinfinal} which differ only by the fixing of three out of six functions are nothing but equivalent representations of the same physical theory. 

It is worth noticing that the invariance under two metric transformations allows the definition of more physically interesting equivalent frames, w.r.t. standard scalar-tensor theories. In fact, we can actually define the following four equivalent frames, all obtained from the action \eqref{Einsteinfinal} with different fixing of the free functions.

\begin{description}
\item[ Jordan Frame] The Jordan frame is defined by the action
\begin{multline}
S_{J}=\int d^{4}x\sqrt{-g}\left[G_{J}(\phi,X)R+\right.\\
\left.-G_{J,X}\left[(\square\phi)^{2}-(\nabla_{\mu}\nabla_{\mu}\phi)^{2}\right]\right.+\\
\left.+K(\phi,X)+G_{3}(\phi,X)\square\phi\right]+S_{m}[g,\psi]\, ,
\label{Jordanframe}
\end{multline}
where we have fixed  $\alpha=1$ and $\beta=1$ so that matter is minimally coupled to the metric that defines the curvature terms appearing in the action. As a consequence a conformal non-minimal coupling term, described by the presence of the function $G_{J}=C(\phi)^{2}(1-2D(\phi)X)^{1/2}$, is present and can be further constrained with a field redefinition.

\item[Einstein Frame] The Einstein frame is given by the action
\be
S_{E}=\int d^{4}x\sqrt{-g}\left[R+K(\phi,X)+G_{3}(\phi,X)\square\phi\right]+S_{m}[\bar g,\psi]\, ,
\label{Einsteinframe}
\ee
where the NMC has been eliminated by the fixing $C(\phi)=1$ and $D(\phi=0)$ in the action (\ref{Einsteinfinal}) but now matter feels a physical metric related via a disformal transformation to that defining curvature terms, \ie $ \bar g_{\mu\nu}=e^{\alpha(\phi)}g_{\mu\nu}+\beta(\phi)\phi_{\mu}\phi_{\nu}$. Again we can fix one of the two functions $\alpha$ and $\beta$ via a field rescaling.

\item[Galileon Frame] This frame is given by the action
\begin{multline}
S_{G}=\int d^{4}x\sqrt{-g}\left[G_{G}(\phi,X)R-G_{G,X}\left[(\square\phi)^{2}-(\nabla_{\mu}\nabla_{\mu}\phi)^{2}\right]\right.\\
\left.+K(\phi,X)+G_{3}(\phi,X)\square\phi\right]+S_{m}[\bar g,\psi]\, ,
\label{Galileonframe}
\end{multline}
where
\be
G_{G}= (1-2D(\phi)X)^{1/2};\qquad \bar g_{\mu\nu}=e^{\alpha(\phi)}g_{\mu\nu}\, ,
\ee
which amounts to the choice $C(\phi)=1$ and $\beta(\phi)=1$. In this case we have both NMC and matter fields feeling a physical metric which is now conformally related to the gravitational one. 

\item[Disformal Frame]This frame is given by the action
\begin{multline}
S_{D}=\int d^{4}x\sqrt{-g}\left[G_{G}(\phi,X)R+K(\phi,X)+\right.\\
\left.G_{3}(\phi,X)\square\phi\right]+S_{m}[\bar g,\psi]\, ,
\label{Disformalframe}
\end{multline}
where
\be
G_{G}= C(\phi)^{2};\qquad \bar g_{\mu\nu}=g_{\mu\nu}+\beta(\phi)\phi_{\mu}\phi_{\nu}\, ,
\ee
which amounts to the choice $D=0$ and $\alpha=1$.

\end{description}

It is worth stressing here that the last two frames, which can be seen as some sort of intermediate frames between the Jordan and Einstein ones, can actually reduce to the latter for suitable choices of the rescaling of the field, as can be seen from the last columns of table \ref{tab:frames}. This is a consequence of the fact that we have four free metric functions, $\alpha(\phi)$, $\beta(\phi)$, $C(\phi)$ and $D(\phi)$, three of which can be arbitrarily fixed and hence some overlap is expected. 
 
As a final remark, while all these equivalent frames are connected by disformal transformations and field rescaling, one has also to be careful about accordingly rescale also the so far neglected functions $K(\phi,X)$ and $G_{i}(\phi,X)$ in order to preserve the equivalence of frames.

The above mentioned frames where first proposed in \cite{Mota2013} and partially discussed in \cite{Rham} where it was pointed out how disformal transformations relate  them, albeit no discussion about their actual equivalence was provided. Here we have re-derived the same results in a different way and in addition we have proved the frames equivalence. This has relevant consequences, for example it implies that not only DBI Galileon models with a non-minimally coupled scalar field can be cast via a disformal transformation into the simpler Einstein frame, but also guarantees the equivalence of these representation.  Furthermore, the equivalence of the frames allows us to claim the equivalence of many apparently unrelated models as those reported in \cite{Zumalacarregui2010} given that we can move from one to the other through appropriately chosen disformal transformations and field redefinitions.
\begin{widetext}
\begin{center}
\begin{table}
\centering
{
\begin{tabular}{ccccc}

   		{\bf Frame}	      &		    	\multicolumn{2}{c}{\bf{Disformal transformation}}		       & \multicolumn{2}{c}{\bf{Field rescaling}}   \\
					      &	 {\bf Matter Metric}	    &		 {\bf NMC function 	}			       &{\bf Matter Metric}  &{\bf  NMC function} \\
\hline
 \multirow{2}*{ Jordan Frame}& \multirow{2}*{$g_{\mu\nu}$}&\multirow{2}*{$C(\phi)^{2}(1+2D(\phi)X)^{1/2}$}&  $g_{\mu\nu}$&$(1-2D(\phi)X)^{1/2}$\\
 					      &					    &									       &  $g_{\mu\nu}$&$ C(\phi)^{2}(1-2\Lambda X)^{1/2}$\\
					      &					    &									       &			& \\
 \multirow{2}*{ Einstein Frame}& \multirow{2}*{ $e^{\alpha(\phi)}g_{\mu\nu}+\beta(\phi)\phi_{\mu}\phi_{\nu}$}&\multirow{2}*{$1$}& $\varphi(\phi) g_{\mu\nu}+\beta(\phi)\phi_{\mu}\phi_{\nu}$    &$1$\\
 						&														       & 			      &$e^{\alpha(\phi)}g_{\mu\nu}+\Lambda\phi_{\mu}\phi_{\nu}$&$1$\\
					      &					    &									       &			& \\
 \multirow{2}*{ Galileon Frame}& \multirow{2}*{$e^{\alpha(\phi)}g_{\mu\nu}$ }&\multirow{2}*{$(1-2D(\phi)X)^{1/2}$}& $ \varphi(\phi)g_{\mu\nu}$     &$(1-2D(\phi)X)^{1/2}$\\
 						& 							       &							 &$e^{\alpha(\phi)}g_{\mu\nu}$&$(1-2\Lambda X)^{1/2}$\\
											      &					    &									       &			& \\
 \multirow{2}*{ Disformal Frame}& \multirow{2}*{$g_{\mu\nu}+\beta(\phi)\phi_{\mu}\phi_{\nu}$}&\multirow{2}*{$C(\phi)^{2}$}& $g_{\mu\nu}+\Lambda \phi_{\mu}\phi_{\nu}$&$ C(\phi)^{2}$\\
 & & &$g_{\mu\nu}+\beta(\phi)\phi_{\mu}\phi_{\nu}$ &$1$\\
  \hline
\end{tabular}
}
\caption{Disformal frames obtained for different fixing of the Horndeski coefficient functions of \eqref{Einsteinfinal}. The first two columns show the results of the fixing after a disformal transformation while the last two  show the effects of the further freedom associated to the invariance under field rescaling (there are two possibilities in each slot in this case as one can alternatively rescale the metric or the field $\phi$ derivative terms). $\Lambda$ is a dimensional constant introduced to keep track of the dimensions of the coefficient functions, while $\varphi(\phi)$ is the rescaled conformal function.}
\label{tab:frames}
\end{table}
\end{center}
\end{widetext}

\subsection{More general disformal frames}

We have seen in the previous section that the requirement of an Einstein frame strongly constrains the shape of the Horndeski Lagrangian with a specific form for $G_{4}(\phi,X)$ and forcing $G_{5}(\phi,X)=0$. However, there is no real physical need to have an Einstein frame so that one may wonder about the existence of more general Lagrangians that do not posses an Einstein frame but that show in any case interesting properties under disformal transformation. We list and analyse here some examples.
\begin{description}
\item[Disformal matter] When we add the matter Lagrangian to the full Horndeski action, the EEP allows matter fields to be coupled to a metric which is disformally related to the one defining the Horndeski action
\be
S= S_{H}[g,\phi]+S_{m}[\bar g,\psi]\,,
\ee
where $S_{H}$ is the full Horndeski action \eqref{Horndeski}, $\psi$ collectively defines matter fields and where $\bar g_{\mu\nu}=e^{\alpha(\phi)}g_{\mu\nu}+ \beta(\phi)\phi_{\mu}\phi_{\nu}$. Thanks to the invariance of the full Horndeski action under disformal transformations and field rescaling we are free to fix both $\alpha(\phi)$ and $\beta(\phi)$ in such a way that, after the transformation, matter propagates along the geodesics defined by the metric $g_{\mu\nu}$ that appears in the Horndeski action. These transformations will of course affect the Horndeski Lagrangian, but only in the shape if its coefficient functions, not in its structure. 
Hence, a Horndeski theory in which matter propagates on the metric $\bar g_{\mu\nu}=e^{\alpha(\phi)}g_{\mu\nu}+ \beta(\phi)\phi_{\mu}\phi_{\nu}$ is equivalent to another Horndeski theory, with redefined coefficient functions, in which matter propagates along the same metric $g_{\mu\nu}$ that enters the Horndeski action.

This fact is not particularly surprising but it is nonetheless interesting as it shows how, without any assumption on the shape of the Horndeski action, we can see that apparently different matter behaviours are in fact different representations of the same theory. 

\item[Einstein coupling] Another possible extension is to include $\mtc{L}_{5}$ Lagrangian while keeping the requirement of having a frame with no conformal coupling to gravity. Using the relations derived in appendix (\ref{disftrans}) we see that this requirements translate into a condition on the initial shape of the $G_{4}(\phi,X)$ function
\be
G_{4}(\phi,X)=(1-2B(\phi)X)^{1/2}-G_{5,\phi}(\phi,X)X +\tilde G_{5,\phi}(\phi,X)X\, ,
\ee
where, 
\be
\tilde G_{5}(\phi,X)=\int dX (1-2BX)^{1/2}G_{5,X}(\phi,X)\, .
\ee
With this requirement we can consider the following action
 \begin{multline}
S=\int d^{4}x\sqrt{-g}\left[G_{4}(\phi,X)R-G_{4,X}(\phi,X)\left[(\square\phi)^{2}\right.\right.\\
\left.\left.-(\nabla_{\mu}\nabla_{\mu}\phi)^{2}\right]+K(\phi,X)+G_{3}(\phi,X)\square\phi\right]+\\
+\int d^{4}x\sqrt{-g}\left[G_{5}(\phi,X)G_{\mu\nu}\nabla^{\mu}\nabla^{\nu}\phi -\frac{1}{6}\left((\square\phi)^{3}\right.\right.\\
\left.\left.-3\square\phi(\nabla_{\mu}\nabla_{\nu}\phi)^{2}+2(\nabla_{\mu}\nabla_{\nu}\phi)^{3}\right)\right]+S_{m}[\tilde g,\psi]\, ,
\label{EinsteinEinstein}
\end{multline}
where $G_{4}(\phi,X)$ is given by the previous expressions while $G_{5}(\phi,X)$ is left totally free. With the disformal transformation (\ref{disformal}) we can eliminate the conformal coupling to gravity leaving only a NMC via the Einstein tensor and matter fields propagating along disformal geodesics.

\end{description}

We conclude this section recalling that the invariance of the Horndeski action under disformal transformations and field rescaling holds true for the full Horndeski theory \eqref{Horndeski}. Possible restrictions on the shape and functional dependencies of the free functions of the theory are to be ascribed only to physical motivations, \eg the requirement of an Einstein frame, or to classification aims, \eg identify equivalent models, but not to constraints imposed by the invariance itself.

\section{Conclusions}
\label{conclusions}

The gravitational interaction has been the first one studied in a systematic way and its modern formulation is encoded in the theory of General Relativity. Despite its successes, GR is nowadays challenged both at the theoretical and experimental level, leading to several proposals for alternatives theories of gravity. However the lack of an axiomatic procedure for the construction of such theories and the limited regime for which we have highly constraining observational data, make hard to reduce the number of alternative theories and to find their mutual relations.

A major tool in physics is represented by symmetries. This is a clean and precise way to order models, find their simplest formulations and identify the minimal set of degrees of freedom required to fully define a theory. In the context of standard scalar-tensor theories this has been systematically investigated and the discovery of the invariance of such theories under conformal metric transformations and field rescaling has made possible to identify the minimal number of functions required to describe the theory and showed the mutual relations between apparently different representations.

Along this line of reasoning, we have investigated in this paper the symmetries of the Horndeski action and found that it is invariant under a more general metric transformation than conformal, the so called disformal transformation, as well as under field rescalings. These transformations contain free functions and hence can in principle be used to constrain the coefficient functions that define the Horndeski action. However, we have shown that the most general disformal transformation \eqref{generaldisformal} cannot be used to this purpose as the Horndeski action is not invariant under transformations induced by it. We have hence circumscribed our investigation to a subset of disformal transformations, \eqref{disformal} where the two free functions needed to define it only depends on the scalar field. We have shown that the Horndeski action is actually invariant under such class of disformal transformations albeit the generality of the Horndeski action does not allow for an efficient fixing of the coefficient functions. 

For this reason, we looked to the constraints that one has to impose on the Horndeski coefficient functions in order to have a theory that admits an Einstein frame. We discovered that this is a quite constraining request as in fact the full Horndeski action is reduced to the action \eqref{Einsteinfinal} where only a conformal non-minimal coupling is present. This allowed us to investigate the existence of equivalent frames, in an analogous way as what is done for standard scalar-tensor theories. We found that apart from the well known Einstein and Jordan frames, the invariance under disformal transformations allows for the definition of two more equivalent frames: the so called Galileon and Disformal frames. We further extend our analysis to frames that do not admit an Einstein frame and showed that even without this requirement one can found physically relevant frames connected by disformal transformations.

In conclusions, with this work we have found a new class of scalar-tensor theories of gravity that admits disformally equivalent frames, which are related by disformal transformation and field rescaling, thus generalizing the previous results obtained in the context of standard scalar-tensor theories. This may have important consequences in cosmological context as may allow to identify a large class of models into different representations of the same theory. 
We hope that this issues will be further investigated in a near future.

\acknowledgments
The Authors wish to thank Thomas Sotiriou and Daniele Vernieri for discussions and Miguel Zumalacarregui for useful comments. Our calculations have been cross-checked with xTensor package \cite{xtensor} developed by J.-M. Martin-Garcia for Mathematica.

\appendix

\section{Keeping second order field equations}
\label{NoX}

In this section we show how a metric transformation induced by the general disformal relation \eqref{generaldisformal} spoils the property of the Horndeski action of producing second order field equations.\footnote{ We want to stress that this result holds on both curved background as well as on flat backgrounds with the exception that on flat space times there exist subcases that give second order field equations even after a disformal transformation.}

Our proof consists in a direct calculation of the modifications that the disformal transformation has onto a particular term of the full Lagrangian, namely $\mtc{L}_{4}$, when the disformal functions depends only on the kinetic term of the scalar field $\phi$. Despite this does not represent a formal proof of our statement it is nonetheless general enough to discard any kinetic term dependence in the disformal transformation if second order field equations are to be preserved. We leave the formal proof of this for further work, but we stress that the result obtained here holds in general. Our calculations make use of \cite{Deffayet2011}, where a general procedure on how to build actions for a metric and a scalar field that keeps the equation of motion second order was put forward. We will shortly review it for what concerns us, referring the interested reader to the original paper.\footnote{Notice that in our work we have the following correspondences: $\pi\rightarrow\phi$, $\pi_{\mu}\rightarrow\phi_{\mu}$, $\pi_{\mu\nu}\rightarrow\nabla_{\mu}\nabla_{\nu}\phi$ and $X\rightarrow2X$.}

In flat space times consider the following Lagrangian:
\be
\mathcal{L}=\mathcal{T}_{(2n)}^{\mu_{1}\cdots\mu_{n}\nu_{1}\cdots\nu_{n}}\nabla_{\mu_{1}}\nabla_{\nu_{1}}\phi\cdots\nabla_{\mu_{n}}\nabla_{\nu_{n}}\phi\,,
\label{Ldef}
\ee
where
\be
\mathcal{T}=\mathcal{T}(\phi,\phi_{\alpha}), \qquad \mathcal{L}=\mathcal{L}(\phi,\phi_{\mu},\nabla_{\mu}\nabla_{\nu}\phi)\,,
\ee
then the following lemma holds:
\begin{Lemma}
A sufficient condition for the field equations derived from the Lagrangian \ref{Ldef} to remain second order or less is that $\mathcal{T}_{(2n)}^{\mu_{1}\cdots\mu_{n}\nu_{1}\cdot\nu_{n}}$ is totally antisymmetric in its first indices $\mu_{i}$ as well as (separately) in its last indices $\nu_{i}$.
\end{Lemma}

Notice that this is a sufficient conditions. However the opposite statement has been proven and a uniqueness condition exists so that the condition is both necessary and sufficient.

When one moves to curved space times and covariantize promoting partial derivatives to covariant derivatives third order derivatives of the metric are produced. It has been shown that adding a suitable finite number of non-minimally coupled terms to the Lagrangian is enough to eliminate the higher than second derivatives from the equations of motion in both the scalar field and in the metric.
As a final result the authors of \cite{Deffayet2011} gave the form of the Lagrangian that preserves the second order equations:
\be
\mathcal{L}_{n}\{f\}=\sum_{p=0}^{\lfloor n/2\rfloor}\mathcal{C}_{n,p}\mathcal{L}_{n,p}\{f\}\,,
\ee
where $\lfloor n/2\rfloor$ indicates the integer part while the graph bracket indicates that $\mathcal{L}$ is a functional of $f$, which is in general different for any $n$, and where
\be
\mathcal{L}_{n,p}\{f\}=P_{(p)}^{\mu_{1}\cdots\mu_{n}\nu_{1}\cdots\nu_{n}}\nabla_{\mu_{1}}R_{(p)}S_{(q\equiv n-2p)}\,,
\ee
\be
R_{(p)}=\prod_{i=1}^{p}R_{\mu_{2i-1}\mu_{2i}\nu_{2i-1}\nu_{2i}}\,,
\ee
\be
S_{(q\equiv n-2p)}=\prod_{i=0}^{q-1}\nabla_{\mu_{n-i}}\nabla_{\nu_{n-i}}\phi\,,
\ee
while
\be
P_{(p)}^{\mu_{1}\cdots\mu_{n}\nu_{1}\cdots\nu_{n}}=\int_{X_{0}}^{X}dX_{1}\cdots\int_{X_{0}}^{X_{p-1}}dX_{p}\mathcal{T}_{(2n)}^{\mu_{1}\cdots\mu_{n}\nu_{1}\cdots\nu_{n}}(\phi,X_{1})\,,
\ee
while the coefficients are given by:
\be
\mathcal{C}_{n,p}=\left(-\frac{1}{8}\right)^{p}\frac{n!}{(n-2p)!p!}\,.
\ee

Using the Lagrangian \ref{Ldef} and the rules reported above it is possible to construct all covariant theories that gives second order field equations and in particular in four dimensions we have that the Horndeski action is a linear combination of the following terms:
\be
\mathcal{L}_{0,0}=Xf_{0}(\phi,X)\,,\qquad
\mathcal{C}_{0,0}=1\,,
\ee
\be
\mathcal{L}_{1,0}=Xf_{1}(\phi,X)A_{2}^{\mu\nu}\nabla_{\mu}\nabla_{\nu}\phi\,, \qquad
\mathcal{C}_{1,0}=1\,,
\ee
\be
\mathcal{L}_{2,0}=Xf_{2}(\phi,X)A_{4}^{\mu_{1}\mu_{2}\nu_{1}\nu_{2}}\nabla_{\mu_{1}}\nabla_{\nu_{2}}\phi\nabla_{\mu_{2}}\nabla_{\nu_{2}}\phi\,,\quad
\mathcal{C}_{2,0}=1\,,
\ee
\be
\mathcal{L}_{3,0}=Xf_{3}(\phi,X)A_{6}^{\mu_{1}\mu_{2}\mu_{3}\nu_{1}\nu_{2}\nu_{3}}\nabla_{\mu_{1}}\nabla_{\nu_{2}}\phi\nabla_{\mu_{2}}\nabla_{\nu_{2}}\phi\nabla_{\mu_{3}}\nabla_{\nu_{3}}\phi\,,\qquad
\mathcal{C}_{3,0}=1\,,
\ee
\begin{multline}
\mathcal{L}_{2,1}=P_{(1)}^{\mu_{1}\mu_{2}\nu_{1}\nu_{2}}R_{\mu_{1}\mu_{2}\nu_{1}\nu_{2}}\,,\\ P_{(1)}=\int dX_{1}\mathcal{A}_{4}^{\mu_{1}\mu_{2}\nu_{1}\nu_{2}}X_{1}f_{(2)}(\phi,X_{1})\,,
\quad\mathcal{C}_{2,1}=-\frac{1}{4}\,,
\end{multline}
\begin{multline}
\mathcal{L}_{3,1}=P_{(1)}^{\mu_{1}\mu_{2}\mu_{3}\nu_{1}\nu_{2}\nu_{3}}R_{\mu_{1}\mu_{2}\nu_{1}\nu_{2}}\nabla_{\mu_{3}}\nabla_{\nu_{3}}\phi\,,\\ P_{(1)}=\int dX_{1}\mathcal{A}_{6}^{\mu_{1}\mu_{2}\mu_{3}\nu_{1}\nu_{2}\nu_{3}}X_{1}f_{(3)}(\phi,X_{1})\,,\quad
\mathcal{C}_{3,1}=-\frac{3}{4}\,,
\label{rephrased}
\end{multline}

where we have redefined the form function $\mathcal{T}_{2n}(\phi,X)=X f_{n}(\phi,2X)\mathcal{A}_{2n}$ in such a way to separate the field dependences $(\phi,X)$ from the structure term $\mtc{A}(g_{\alpha\beta},\phi_{\alpha})$. Notice that the terms $(2,0)$ and $(2,1)$ as well as $(3,0)$ and $(3,1)$ are coupled terms whose joint presence is required in order to cancel the unwanted higher order derivatives.
The Horndeski action can be rephrased in these terms with the following identifications:\footnote{Notice that compared with the convention used in the definition of the kinetic term X in \cite{Deffayet2011} there are factors 1/2 that have been reabsorbed into the definition of the function $f_{(n)}$.}
\ba
K(\phi,X)&=&Xf_{(0)}(\phi,X)\,, \\
G_{3}(\phi,X)&=&Xf_{(1)}(\phi,X)\, ,\quad \mtc{A}_{(2)}^{\mu\nu}=g^{\mu\nu}\, ,
\ea
\ba
G_{4}(\phi,X)&=&\int\left[X_{1}f_{(2)}(\phi,X_{1})dX_{1}\right]\,,\\
 \mtc{A}_{(4)}^{\mu\alpha\nu\beta}&=&g^{\mu\beta}g^{\nu\alpha}-g^{\mu\nu}g^{\alpha\beta}\, ,
\ea
\be
G_{5}(\phi,X)=\int\left[X_{1}f_{(3)}(\phi,X_{1})dX_{1}\right]\,,
\ee
\begin{multline}
 \mtc{A}_{(6)}^{\mu\sigma\alpha\nu\rho\beta}=g^{\alpha\nu}\left[g^{\beta\mu}g^{\sigma\rho}-g^{\beta\rho}g^{\sigma\mu}\right]\\
 +g^{\alpha\sigma}\left[-g^{\beta\mu}g^{\nu\rho}+g^{\beta\rho}g^{\nu\mu}\right]+g^{\alpha\beta}\left[g^{\sigma\mu}g^{\nu\rho}-g^{\sigma\rho}g^{\nu\mu}\right]\,.
\end{multline}
In order to prove our statement we need to show how a kinetic dependent metric transformation spoils the antisymmetric structure of the model. Before entering the calculations we note that the effects of a disformal metric transformation on the coefficient functions $f_{(n)}$ only redefines its functional dependence, while the structure functions $\mtc{A}$ are again only redefined with no modifications on their antisymmetric structure. Hence, in order to check the breaking of the antisymmetric structure, we only need to compute the effects of metric transformations on second covariant derivatives of the field and on the Riemann tensor.
In order to do this in a simple way, we will look at the effects of the kinetic dependence of the disformal functions $A(X)$ and $B(X)$ by applying separately a conformal transformation and a purely disformal one on the terms corresponding to $\mathcal{L}_{4}$ in the rephrased Horndeski action \eqref{rephrased}.  

\subsection{Conformal transformation}

Consider a conformal transformation of the kind
\be
\bar{g}_{\mu\nu}=A(X)g_{\mu\nu}.
\label{genconf}
\ee

After the conformal transformation \ref{genconf} is performed the original $\mathcal{L}_{4}$ Lagrangian is mapped into:
\begin{multline}
A^{2}G_{4}R-A^{2}G_{4,X}\left[(\square\phi)^{2}-(\nabla_{\mu}\nabla_{\nu}\phi)^{2}\right]\\
-6AA'G_{4}(\nabla_{\mu}\nabla_{\nu}\phi)^{2}-6AA'G_{4}\phi^{\alpha}\square\nabla_{\alpha}\phi\\
-2AA'G_{4,X}\left[\square\phi+\frac{A'}{A}\phi^{\mu}\phi^{\nu}\nabla_{\mu}\nabla_{\nu}\phi\right]\phi^{\alpha}\phi^{\beta}\nabla_{\alpha}\nabla_{\beta}\phi\\
+\phi^{\mu}\phi^{\nu}\nabla_{\mu}\nabla_{\alpha}\phi\nabla_{\nu}\nabla^{\alpha}\phi\left[-4AA'G_{4,X}+AA'^{2}X-6A''AG_{4}\right]\,.
\end{multline}
From this expression it is clear that the first two terms are not dangerous as they have the same structure as those in the original Lagrangian. In order to better understand the others we proceed in rewriting them in the form $\mtc{A}^{\mu\alpha\nu\beta}\nabla_{\mu}\nabla_{\nu}\phi\nabla_{\alpha}\nabla_{\beta}\phi$. Any antisymmetry violating term will then directly lead to higher derivatives in the equation of motions. After some manipulation we arrive at the expression:
\begin{multline}
\sim \left[-6AA'G_{4}(g^{\alpha\mu}g^{\beta\nu})\right.
\\ \left.+(-2G_{4,X}AA'+6A'^{2}G_{4}+AA''G_{4}+AA'G_{4,X})g^{\mu\nu}\phi^{\alpha}\phi^{\beta}+\right.\\
\left.+(-4G_{4,X}AA'+4G_{4,X}A'^{2}X-6AA''G_{4})g^{\nu\beta}\phi^{\alpha}\phi^{\mu}\right.\\
\left.-2G_{4,X}A'^{2}\phi^{\mu}\phi^{\nu}\phi^{\alpha}\phi^{\beta}\right]\nabla_{\mu}\nabla_{\nu}\phi\nabla_{\alpha}\nabla_{\beta}\phi\, ,
\end{multline}
where the symbol $	\sim$ indicates that only the dangerous terms have been considered and notice that we have added a surface terms to rewrite the third order derivative.  As can be easily seen antisymmetry breaking terms have appeared in the Lagrangian.
We can then conclude that the generalised conformal transformation \ref{genconf} spoils the antisymmetric structure of the Horndeski action and hence gives equation of motion for the fields that are higher than second order. 

\subsection{Disformal transformations}

Consider now a metric transformation of the form
\be
\bar{g}_{\mu\nu}=g_{\mu\nu}+B(X)\phi_{\mu}\phi_{\nu}\, .
\ee
Using the same procedure of the previous section we can write the transformed $\mtc{L}_{4}$ part of the Lagrangian and see whether or not it is possible to recover the antisymmetric structure. The dangerous terms of the transformed Lagrangian read
\begin{widetext}
\begin{multline}
\sim\left[g^{\mu\nu}\phi^{\alpha}\phi^{\beta}\left(\frac{2G_{4,X}}{\fdB^{1/2}}(B'X+B)-\frac{2G_{4}}{\fdB^{3/2}}(B^{2}-B'(1+BX))-2\frac{G_{4}}{\fdB^{1/2}}R_{\alpha\mu\beta\nu}\right.\right.\\
\left.\left.-\frac{G_{4}B'}{\fdB^{1/2}}+(\frac{2B'G_{4}}{\fdB^{1/2}}+\frac{2XB''G_{4}}{\fdB^{1/2}}+\frac{2B'G_{4,X}}{\fdB^{1/2}}-\frac{2B'G_{4}}{\fdB^{3/2}}(B'X+B)\right)\right]+\\
g^{\mu\alpha}\phi^{\beta}\phi^{\nu}\left[-\frac{G_{4,X}}{\fdB^{1/2}}(B-2XB'(-1X^{2}B'))+\frac{G_{4}}{\fdB^{3/2}}(B^{2}-B'+B'^{2}X^{2}-XB''\fdB)\right.\\
\left.+\frac{G_{4}B'}{\fdB^{1/2}}-2\frac{G_{4,X}B'X}{\fdB^{1/2}}-2\frac{G_{4}B'}{\fdB^{1/2}}\right]+
\phi^{\mu}\phi^{\nu}\phi^{\alpha}\phi^{\beta}\left[\frac{G_{4,X}B'}{\fdB^{1/2}}(1-2X^{2}B')+\right.\\
\left.\frac{G_{4}}{\fdB^{3/2}}(-XB'^{2}+B''\fdB-BB')-\frac{G_{4,X}B'}{\fdB^{1/2}}\right.\\
\left.-\frac{G_{4}B''}{\fdB^{1/2}}+\frac{G_{4}B'}{\fdB^{3/2}}(B'X+B)\right]\nabla_{\mu}\nabla_{\nu}\nabla_{\alpha}\phi\nabla_{\beta}\phi\, ,
\end{multline}
\end{widetext}
which, again, contains terms which are not antisymmetric in the couples $(\alpha,\beta)$ and $(\mu,\nu)$ hence giving rise to higher derivatives in the equations of motion.
\newline

In conclusion, even if a formal proof of this result would be desirable, our result clearly states that if one wants to preserve second order field equations, then the most general disformal transformation that can be used is the one reported in eq. \ref{disformal} where the disformal functions $A$ and $B$ only depends on the scalar field $\phi$.

\section{Transformation properties of geometrical quantities.}
\label{transprop}

We provide here the transformation rules for geometric quantities when the metric undergoes a disformal transformation of the kind
\be 
\bar{g}_{\mu\nu}=A(\phi)g_{\mu\nu}+B(\phi)\phi\mu\phi\nu\, ,
\label{A:disformal}
\ee
where both metric $g$ and $\bar{g}$ are well defined metrics that can be equally be used to rise and lower indices.
The transformed inverse is:
\be
\bar{g}^{\mu\nu}=\frac{1}{A(\phi)}g^{\mu\nu}-\frac{B(\phi)}{A(\phi)^{2}\fd}\phi^{\mu}\phi^{\nu}\, ,
\ee
while the volume element changes (see appendix C of \cite{Bekenstein2004}) as
$ \sqrt{-\bar{g}}=A(\phi)^{2}\fd^{1/2}\sqrt{-g}$.

From this definitions one can express all the barred curvature quantities in function of the unbarred metric and the scalar field $\phi$. We list these below.
\begin{widetext}
\begin{description}
\item[Connection coefficient]
\begin{multline}
\bar{\Gamma}^{\alpha}_{\phantom{\alpha}\mu\nu}=\Gamma^{\alpha}_{\phantom{\alpha}\mu\nu}+\frac{B}{A\fd}\phi^{\alpha}\nabla_{\mu}\nabla_{\nu}\phi+\frac{A'}{2A}\left(\delta^{\alpha}_{\nu}\phi_{\mu}+\delta^{\alpha}_{\mu}\phi_{\nu}\right)\\
+\frac{1}{2}\frac{\phi^{\alpha}}{A^2\fd}\left(-AA'g_{\mu\nu}+(AB'-2A'B)\phi_{\mu}\phi_{\nu}\right)\, .
\end{multline}

\item[Ricci Tensor]
\begin{multline}
\bar{R}_{\alpha\beta}=R_{\alpha\beta}+\frac{\left[AB\fd \square\phi-B^{2}\phi^{\mu}\nabla_{\mu}X-AA'\fd+(AB'-A'B)X\right]}{A^{2}\fd^{2}}\nabla_{\alpha}\nabla_{\beta}\phi+\\
+\frac{\left[-A^{2}A'\fd\square\phi+AA'B\phi^{\mu}\nabla_{\mu}X-2A'X^{2}(A'B-AB')-2A^{2}A''X\fd\right]}{2A^{3}\fd^{2}}g_{\alpha\beta}+\\
+\left[\frac{3A^{2}A'^{2}+6A'^{2}B^{2}X^{2}+2A'A^{2}B'X+\left(A^{3}B'-4AA'B^{2}X\right)\square\phi-2A^{3}A''}{2A^{4}\fd^{2}}\right.\\
+\left.\frac{2AB\left(A'B\nabla^{\mu}\nabla_{\mu}X+5A'^{2}X+3A'B'X^{2}\right)-6A''BX^{2}}{2A^{4}\fd^{2}}\right]\phi_{\alpha}\phi_{\beta}+\\
+\left[\frac{-2AB\fd R_{\alpha\mu\beta\nu}\phi^{\mu}\phi^{\nu}-2AB\fd\nabla_{\alpha}\nabla_{\lambda}\phi\nabla_{\beta}\nabla^{\lambda}\phi+2B^{2}\nabla_{\alpha}X\nabla^{\alpha}X}{2A^{2}\fd^{2}}\right.\\
+\left.\frac{(A'B-AB')(\phi_{\alpha}\nabla_{\beta}X+\phi_{\beta}\phi_{\alpha})-\phi_{\alpha}\phi_{\beta}\phi^{\mu}\phi\nabla_{\mu}X+2\phi_{\alpha}\phi_{\beta}\phi\square\phi(A'-B'X)+10XA'' \phi_{\alpha}\phi_{\beta}\phi}{2A^{2}\fd^{2}}\right]\, .
\end{multline} 
\item[Ricci Scalar]
\begin{multline}
\bar{R}=R-\frac{2B}{A^{2}\fd}R_{\alpha\beta}\phi^{\alpha}\phi^{\beta}+\frac{B}{A^{2}\fd}\left[(\square\phi)^{2}-(\nabla_{\alpha}\nabla_{\beta}\phi)^{2}\right]+\\
+\frac{2B}{A^{3}\fd^{2}}\left[\nabla^{\alpha}X\nabla_{\alpha}X-\phi^{\alpha}\nabla_{\alpha}X\square\phi\right] -\frac{8A'BX+A(3A'-2B'X)}{A^{3}\fd^{2}}\square\phi+\\
+\frac{4A'B-AB'}{A^{3}\fd^{2}}\phi^{\alpha}\nabla_{\alpha}X
+\frac{3A'X(A'+2B'X)}{A^{2}\fd^{2}}-\frac{6A''X}{A^{2}\fd}\, .
\end{multline}
\end{description}
\end{widetext}
Notice that both functions $A$ and $B$ are to be intended as general functions of the scalar field $\phi$.

\section{Transformation properties of the Horndeski action under disformal transformations}
\label{disftrans}

We explored the consequences on the Horndeski action when the metric is transformed via a disformal transformation
\be 
\bar{g}_{\mu\nu}=A(\phi)g_{\mu\nu}+B(\phi)\phi_{\mu}\phi_{\nu}\, .
\label{app:disformal}
\ee
through a direct calculation. Our results show that after this transformation is performed the new action can be recast into the same initial Horndeski form given that all the effect of the transformation are absorbed into the rescaling of the free coefficient functions. As a consequence we can say that the Horndeski action is formally invariant under this class of disformal transformation.
We report below the transformations properties of the Horndeski Lagrangian coefficient functions. The new Lagrangian is
\be
\mathcal{\bar L}=\sum_i\bar{\mathcal{L}}_i \, ,
\ee
where
\ba
\bar{\mathcal{L}}_2&=&\bar K(\phi,X)\, ,\\
\bar{\mathcal{L}}_3&=&\bar G_3(\phi,X)\square\phi\, ,\\
\nonumber
\bar{\mathcal{L}}_4&=&\bar G_4(\phi,X)R\\
&-&\bar G_{4,X}(\phi,X)\left[(\square\phi)^2-(\nabla_\mu\nabla_\nu\phi)^2\right]\, ,   \\
\nonumber
\bar{\mathcal{L}}_5&=&\bar G_5(\phi,X)G_{\mu\nu}\nabla^\mu\nabla^\nu\phi\\
\nonumber
&+&\frac{\bar G_{5,X}(\phi,X)}{6}\left[(\square\phi)^3-3(\square\phi)(\nabla_\nu\nabla_\mu\phi)^2\right.\\
&+&\left.2(\nabla_\mu\nabla_\nu\phi)^3\right]\, ,
\ea
where
\begin{widetext}
\begin{multline}
\bar K(\phi,X)=\fd^{1/2}K(\phi,X_{D})+2X\left[\frac{G_{3}(\phi,X_{D})AA'}{\fd^{1/2}}+\frac{G_{3}(\phi,X_{D})(A'B)X}{\fd^{3/2}}+H_{3,\phi}(\phi,X)\right]+\\
+3X\frac{G_{4}(\phi,X_{D})\left[A'+2A'B'X-2AA'-4A'BX\right]}{A\fd^{3/2}}
+12X\frac{G_{4,X}(\phi,X_{D})X\left[A'^{2}BX-AA'B'X\right]}{A^{2}\fd^{1/2}}-2XH_{4,\phi}(\phi,X)\\
\frac{3G_{5}(\phi,X_{D})X^{2}A'}{A^{4}\fd^{5/2}}\left[-A'^{2}BX+2A^{2}A''\fd-A(2A'^{2}+3A'B'X)\right]-2XH_{5,\phi}(\phi,X)+\\
\frac{2G_{5,X}(\phi,X_{D})X^{3}}{A^{4}\fd^{3/2}}\left(A'^{3}BX+AA'(A'+3B'X)\right)\, ,
\label{App:newK}
\end{multline}
\begin{multline}
\bar G_{3}(\phi,X)=\left[\frac{AG_{3}(\phi,X_{D})}{\fd^{1/2}}+H_{3}(\phi,X)\right]+\left[\frac{G_{4}(\phi,X_{D})\left(4AA'B+ABB'X+A'B^{2}X\right)}{A^{2}\fd^{3/2}}+\frac{BG_{4,\phi}(\phi,X_{D})}{\fd^{1/2}}+\right.\\
\left.
 \frac{G_{4,X}\left(AA'BX-2A^{2}A'+2A^{2}B'X \right)}{A^{2}\fd^{1/2}}-H_{4}(\phi,X)\right]\\
 +\left[X\left(-2(H_{\square,\phi}(\phi,X)-H_{R,\phi\phi}(\phi,X))+\frac{G_{5}(\phi,X_{D})}{A^{3}\fd^{5/2}}\left(5A'^{2}BX-A\left(\frac{A'^{2}}{2}+6A'B'X\right)\right)+\frac{2G_{5}(\phi,X_{D})}{A\fd^{3/2}}A''\right.\right.\\
\left.\left.\frac{G_{5,X}XA'}{A^{3}\fd^{3/2}}\left(AA'-2A'BX+4AB'X\right)\right)-H_{5}(\phi,X)\right]
\label{App:newG3}\, ,
\end{multline}
\be
\bar G_{4}(\phi,X)=A\fd^{1/2}G_{4}(\phi,X_{D})-\left(\frac{G_{5}(\phi,X_{D})X^{2}}{A^{2}\fd^{3/2}}(A'B-AB')+H_{R,\phi}(\phi,X)X\right)\, ,
\label{App:newG4}
\ee
\be
\bar G_{5}(\phi,X)=\frac{G_{5}(\phi,X_{D})}{\fd^{1/2}}+H_{R}(\phi,X)\, ,
\label{App:newG5}
\ee

where the explicit form of the functions $H_{i}$ are
\be
H_{4}(\phi,X)=\int dX\left[\frac{G_{4}(\phi,X_{D})\left(4AA'B+ABB'X+A'B^{2}X\right)}{A^{2}\fd^{3/2}}\right]\,,\qquad H_{3}(\phi,X)=B\int dX\frac{G_{3}(\phi,X_{D})}{\fd^{3/2}}\, ,
\ee

\begin{multline}
H_{5}(\phi,X)=\int dX\left[H_{\square,\phi}(\phi,X)-H_{R,\phi\phi}(\phi,X)+\frac{G_{5}(\phi,X_{D})}{2A^{3}\fd^{5/2}}\left(-5A'BX-2A^{2}A''\fd+\right.\right.\\
\left.\left.A(5A'^{2}+6A'B'X)\right)+\frac{G_{5,X}(\phi,X_{D})}{A^{2}\fd^{3/2}}\left(-A'BX+2A(A'+B'X)\right)\right]\, ,
\end{multline}
\be
H_{\square}(\phi,X)=G_{5}(\phi,X_{D})\frac{AA'+(AB'-A'B)X}{A^{2}\fd^{3/2}},\qquad H_{R}(\phi,X)=\frac{B}{A}\int dX \frac{G_{5}(\phi,X_{D})}{\fd^{3/2}}\, ,
\ee
\end{widetext}
 while $X_{D}=X/[A(1+2BX/A )] \label{XDisf}$, and, again, the functions $A$ and $B$ depends on the scalar field $\phi$. The most relevant conclusion is that the effect of the disformal transformation on the Horndeski action can be recast into renormalisation of the coefficient functions, exactly as in the case of conformal transformations for standard scalar-tensor theories, which, we stress, are a subcase of our result. Then notice that, if one starts with a only a subset of the Lagrangians, a disformal transformation will in general produce contributions at all sub-Lagrangians in a hierarchical way. Said in other words, the corrections propagate from higher derivatives down to lower derivatives terms. 

\section{Invariance under field rescaling}
\label{fieldinvariance}

Besides the previously analysed invariance under disformal transformation it can be proved that the Horndeski action is also invariant under the rescaling of the scalar field
\be
\phi=s(\psi)\psi\,.
\ee
In fact, the effects of this transformation can be again reabsorbed into redefinitions of the Horndeski coefficient functions which become
\begin{multline}
\bar K(\psi,\bar X)=K(\psi,\bar X)+2YG_{3}(\psi,\bar X)(2s'+\psi s'')+\\
-2YH_{4,\psi}(\psi,\bar X)+2YH_{\square,\psi}\,,
\end{multline}
\begin{multline}
\bar G_{3}(\psi,\bar X)=(s'\psi +s)G_{3}(\psi,\bar X)-\left(4YG_{4,Y}(\psi,\bar X)+\right.\\
\left.-2G_{4}(\psi,\bar X)\right)\frac{2s'+s'' \psi}{s+\psi s'}+2YH_{5,\psi}-H_{\square}\,,
\end{multline}
\be
\bar G_{4}(\psi,\bar X)=G_{4}(\psi,\bar X)-Y(2s'+\psi s'')G_{5}(\psi,\bar X)\, ,
\ee
\be
\bar G_{5}(\psi,\bar X)=(2s'+s''\psi)G_{5}(\psi,\bar X)\,,
\ee
where
\ba
H_{4}(\psi,\bar X)&=&G_{4}(\psi,\bar X)\frac{2s'+s'' \psi}{s+\psi s'}\,,\\
 H_{5}(\psi,\bar X)&=&(2s'+\psi s'')G_{5}\frac{2s'+s'' \psi}{s+\psi s'}\, ,\\
 H_{\square}(\psi,\bar X)&=&\int d\bar X H_{5,\psi}(\psi,\bar X)\,,
\ea
where $\bar X= (s'(\psi)\psi +s(\psi))^{2}Y$, being $Y=\psi^{\mu} \psi_{\mu}/2$, and where a prime denotes the derivative w.r.t. $\psi$.

The field transformation is in principle arbitrary. However, as can be seen from \eg the $\bar G_{3}$ coefficient, infinities may be generated if $ s+\psi s'=0$. This amounts to say that the solution $s(\psi)=\psi^{-1}$ is excluded from the set of admissible rescaling. This fact is in some sense obvious because it is equivalent to the limit of having no scalar field. 
A second remark concerns the possibility to eliminate the NMC with the Einstein tensor with a field redefinition. In fact the transformed $G_{5}$ coefficient is proportional to $2s'(\psi)+\psi s''(\psi)$. This equation can be integrated once giving  $s(\psi)=-\psi s'(\psi)$, whose solution is excluded by the previous requirement. We conclude that it is not possible to eliminate the NMC with the Einstein tensor with a field redefinition.

\end{document}